\documentclass[sigconf]{acmart}
\renewcommand\footnotetextcopyrightpermission[1]{} 
\usepackage{booktabs} 
\usepackage{subfigure}
\usepackage{float}
\usepackage{indentfirst}
\usepackage{listings}
\usepackage[ruled,linesnumbered]{algorithm2e}
\usepackage{amsfonts}

\begin{document}
\title{Interface-Based Side Channel Attack Against Intel SGX}

\author{Jinwen Wang}
\affiliation{%
  \institution{Tsinghua-Berkeley Shenzhen Institute (TBSI), Tsinghua University}
  \city{Shenzhen}
  \state{Guangdong}
  \country{China}
}
\email{wangjw16@mails.tsinghua.edu.cn}

\author{Yueqiang Cheng}
\affiliation{%
  \institution{X-Lab, Baidu}
  \city{Sunnyvale}
  \state{CA}
  \country{USA}
}
\email{strongerwill@gmail.com}


\author{Qi Li}
\affiliation{%
  \institution{Computer Science and Technology, Tsinghua University}
  \city{Shenzhen}
  \state{Guangdong}
  \country{China}
 }
\email{qi.li@sz.tsinghua.edu.cn}

\author{Yong Jiang}
\affiliation{%
  \institution{Tsinghua-Berkeley Shenzhen Institute (TBSI), Computer Science and Technology, Tsinghua University}
  \city{Shenzhen}
  \state{Guangdong}
  \country{China}
}
\email{jiangy@sz.tsinghua.edu.cn}

\renewcommand{\shortauthors}{B. Trovato et al.}

\begin{abstract}
\noindent Intel has introduced a trusted computing technology, Intel Software Guard Extension (SGX), which provides an isolated and secure execution environment called enclave for a user program without trusting any privilege software (e.g., an operating system or a hypervisor) or firmware. Nevertheless, SGX is vulnerable to several side channel attacks (e.g. page-fault-based attack and cache-based attack). In this paper, we explore a new, yet critical side channel attack in SGX, interface-based side channel attack, which can infer the information of the enclave input data. The root cause of the interface-based side channel attack is the input dependent interface invocation information (e.g., interface information and invocation patterns) which can be observed by the untrusted privilege software can reveal the control flow in the enclave. We study the methodology which can be used to conduct the interface-based side channel attack. To illustrate the effectiveness of the interface-based side-channel attacks, we use our methodology to infer whether tracked web pages have been processed by the SGX-assisted NFV platforms and achieve the accuracy of 87.6\% and recall of 76.6\%. We also identify the packets which belong to the tracked web pages, with the accuracy of 67.9\%and recall of 71.1\%. We finally propose some countermeasures to defense the interface-based side channel attack in SGX-assisted applications.   

\end{abstract}

%
%



\keywords{\noindent SGX, Side Channel, NFV, Cloud, Privacy}

\maketitle

\section{Introduction}

Establishing a trusted execution environment (TEE) is one of the most essential security requirement, as we cannot fully trust the underlying computing platforms, such as possible compromised privilege software (e.g. the operating system and hypervisor) and public cloud platforms. Several expressive cryptographic schemes have been proposed to provide trusted computing methods on public cloud platforms. Fully homomorphic or functional encryption\cite{garg2016candidate,smart2010fully,goldwasser2013reusable}  can be used to perform computations directly on the encrypted data. However, these methods are either pretty slow\cite{popa2014building} or basically weak\cite{naveed2015inference, pouliot2016shadow, grubbs2016breaking} . Consequently, new hardware-based primitives, such as ARM TrustZone\cite{ARM}, Trusted Platform Module\cite{Trusted_Computing_Group}, and Intel Software Guard Extension (SGX)\cite{mckeen2013innovative} have been proposed to realize TEEs. Especially, Intel SGX which can provide a secure enclave environment for executing critical codes over sensitive data has received a lot of attention because of its flexibility and availability. Several large Cloud service providers such as Google Cloud\cite{Google.org}, MSR Azure\cite{Azure.org} and IBM Cloud\cite{IBM}, have supported SGX on their Cloud services.

The SGX hardware guarantees that the application memory is secured and any privilege software cannot access the application content. However, SGX-assisted applications rely on the operating system(OS) for system services, such as memory management and scheduling, which will produce side channel that can be exploited to infer the information of enclave data and codes. Several kinds of side channels have been proven to have the ability to leak the information of enclave codes and data, such as cache\cite{brasser2017software}, page-fault\cite{shinde2015preventing}, branch target buffer (BTB) and last branch record (LBR)\cite{lee2017inferring}.

In this paper, we first present a new kind of side channel attack, called the interface-based side channel attack in SGX, which can be exploited to infer the information of enclave input data. The root cause of the interface-based side channel attack is the observable enclave interface (ECALL/OCALL) invocation patterns (e.g. interface parameter sizes and invocation delay etc.) determined by different enclave input data. By collecting and analyzing enclave interface patterns exposed to underlying OS during execution, we can infer the information about enclave input data. Then we will illustrate the methodology which can be used to conduct the interface-based side channel attack, including how to collect the enclave interface invocation patterns in the OS and analyze the collected data. 

To illustrate the effectiveness of the interface-based side channel attack, we try to infer whether tracked website pages have been processed by SGX-assisted Network Function Virtualization (NFV) platforms and identify the packets belonging to the tracked website page on SGX-assisted NFV  platforms\cite{shih2016s,trach2018shieldbox,coughlin2017trusted,duan2017lightbox} which holds the promise of protecting the privacy of processed network traffic. Our experiment result shows that the interface-based side channel attack can identify tracked website page with the accuracy of 87.6\% and recall of 76.6\% and identify the packets which belong to the tracked web pages with the accuracy of 67.9\% and recall of 71.1\%, which breaks the promise of privacy preserving and shows the severity of the interface-based side channel capability. Finally, we proposed some countermeasures which try to eliminate the interface-based side channels and show their effectiveness.

In summary, the contributions of this paper are as follows:

\begin{itemize}
    \item We present a new kind of side channel attack called interface-based side channel attack in SGX which can infer the information of enclave input data by analyzing the observable enclave interface invocation patterns in the OS.
    \item We illustrate the methodology which can be used to conduct the interface-based side channel attack in SGX, including how to collect the enclave interface invocation patterns in the OS and analyze the collected patterns.
    \item To illustrate the effectiveness of interface-based side channel in SGX, we conduct the interface-based side channel attack on SGX-assisted NFV platforms to infer whether tracked website pages have been processed by SGX-assisted NFV platforms and identify the packets which belong to the tracked web site pages.
    \item We finally propose some countermeasures to eliminate the interface-based side channel in SGX and show their effectiveness. 
\end{itemize}

\section{Background}
\subsection{Intel Software Guard Extension}
Intel SGX is a hardware-based mechanism which can ensure the confidentiality and integrity of application code and data even the privilege software or the physical access to the machine (e.g. memory bus and system bus) have been compromised. It depends on two main mechanisms to provide the security promise. One is software isolation mechanism which allows applications to have private data in memory that cannot be accessed by another process, even in the face of system privilege, cause the access is enforced by the processor. Another is remote attestation mechanism which allows a remote verifier to verify whether an enclave has been established on a SGX-enabled system and the integrity of the codes running inside the enclave.

Software isolation mechanism\cite{mckeen2013innovative} is achieved by the new instructions provided by Intel. These instructions can be used to create an  "enclave" which is a memory region that can only be accessed by the creator process. Each enclave is mapped to the enclave page cache (EPC) which is a hardware encrypted address space in main memory access controlled by the processor. The content of EPC is only decrypted inside the processor with processor-specific keys. Thus, even the privilege software (e.g., OS and hypervisor) cannot access the enclave content. The enclave code always executes in user mode, thus any interaction with the OS through the system calls (e.g. network or disk I/O) must execute outside of the enclave.

SGX-assisted application developers can create enclave libraries which can be loaded into an enclave and executed by a CPU supported by SGX through using Intel's SGX SDK\cite{SGX_SDK}. Intel SGX SDK provides a function call mechanism for SGX application by outside call (OCALL) and enclave entry call (ECALL). A developer needs to define the interfaces between the enclave code and other untrusted application code. Specifically, a call into the enclave is referred to as an ECALL, and OCALL allows enclave codes to call untrusted functions outside. The SDK adds instructions to marshal parameters outside the enclave and unmarshal the parameters and execute the function inside the enclave for each ECALL. For each OCALL, the added SDK codes exit the enclave, unmarshals the parameters, and execute the untrusted codes outside the enclave and re-enter the enclave. 

Remote attestation mechanism\cite{anati2013innovative} is achieved by using a challenge-response protocol to generate a measurement of an enclave which is signed by the processor. The measurement can be verified by interacting directly with Intel.
 
By combining the software isolation mechanism and remote attestation mechanism, a remote party can verify the expected codes are running in the enclave before receiving the privacy data. After receiving the secrets, the secrets cannot be accessed by any other part of the remote system.

However, SGX cannot defend any side channel attacks. Especially,  the Intel SGX libraries run outside the enclave. Thus any enclave invocation patterns can be observed by the OS, which can be exploited to infer the information about enclave input data.

\section{Attack}
\subsection{Threat Model}

Our thread model is based on the original threat model of Intel SGX, which assumes the attacker can compromise the OS and exploit the OS to attack the SGX-assisted applications.

First, the attacker can install user level hooks into the untrusted part of a running application which uses enclave interfaces for trusted operation or into the Intel SGX SDK to collect the information about the enclave interface invocation patterns. 

Second, the attacker can control the unit of data sent into the enclave once a time. This is reasonable because the application will process data in a smaller unit when there is not enough data for batch processing, which can help us collect fine-grained enclave interface invocation patterns, and thus reduce the noises. 

Third, the attacker can configure boot properties of the machine, such as running the victim codes on an isolated core, reducing the CPU running frequency and eliminate the eliminable interruptions, which help us measure the enclave interface invocation delay more precise.

Fourth, the attacker can analyze a target enclave program's source codes and/or binary codes in detail to obtain the required information. Unobservable code (e.g, self-modifying code) is outside the scope of our attack.

\subsection{Attack Overview}
In this section, we illustrate the overview of the interface-based side channel attack against the Intel SGX. Our attack mainly includes online phrase and offline phrase. The online phrase is responsible for \textit{interface invocation pattern collection of training data} and \textit{algorithm design and training}. And the offline phrase is responsible for input data inferring. 

In the pattern collection part of the offline phrase, an attack should collect interface invocation patterns for each input data when they are processed by an SGX-assisted application. For a dataset \textit{D}, the attack construct a profiling dataset \textit{P}, each data in \textit{P} can be represented as a vector $d=[f_{1},f_{2},...,f_{n}]$. $f_{i}$ represents the interface-based side channel information collected by an attacker when the data $d$ is processed by an SGX-assisted application. The $f_{i}$ needs to be selected carefully, which can represent useful information about enclave input data in different scenarios.

In the algorithm design part of the offline phrase, we need to design or select algorithms \textit{A} which can use the profiling set to infer the information about new input data information. According to different scenarios, we may need to design different algorithms, which has been well studied in the fields of machine learning. For example, to train the input data with label, we may need to choose supervised learning algorithms, such as k-Nearest Neighbor (KNN) or Decision Tree. To train the input data without label, we may need to select unsupervised learning algorithms, such as Convolutional Neural Network (CNN) or Generative Adversarial Nets (GAN).

In online phrase, we collect the interface invocation patterns during the processing of data $d^{'}=[f_{1}^{'},f_{2}^{'},...,f_{n}^{'}]$. And feed the data $d^{'}$ to the trained model \textit{A} to infer the information about enclave input data.

However, to conduct the interface-based side channel attack successfully, we need to overcome some challenges. First, we need to know all kinds of interface-based side channel information exposed by the SGX-assisted applications, which reveal the enclave input data information. Second, we need to collect these interface-based side channel information precisely.

\subsection{Interface-Based Side Channels}
The Intel SGX libraries are responsible for routines of invoking ECALLs from untrusted codes and routines of invoking OCAlls inside of the enclave, which is under control of the OS. Thus the OS can collect the information about the enclave interface invocation patterns. These information includes the names of ECALLs/OCALLs which are invoked when processing the input data, the size of the parameters in the ECALLs/OCALLs and the processing delay between the different ECALLs/OCALLs invocations. Combining with these information together, the attacker can infer the information about enclave input data, which we call it the interface-based side channel attack against Intel SGX. Specifically, there are three kinds of interface-based side channel information which can be collected by the OS and be exploited to conduct the interface-based side channel attack against the SGX.

\subsubsection{interface invocation sequence}
An application may invoke various ECALLs/OCALLs during execution. Different enclave input data may cause different control flow in the program, which displays different interface invocation sequences. 

For example, the firewall needs to record some event logs when there is traffic from the blocked addresses, which will invoke system calls outside the enclave, such as \textit{write()}. By observing whether OCALLs invoke \textit{write()} when SGX-assisted firewall processing packet, an attacker can infer whether the IP of the processed packet falls into the firewall's block list.

Also, some platforms need different SGX-assisted applications to cooperate to finish tasks. These SGX-assisted applications will execute various ECALLs/OCALLs which will interact with each other. Different input data may cause different ECALL/OCALL invocation sequences, which can be observed by the attacker to infer the input data information. For example, the NFV platform which is used to process web traffic always contains service function chain\cite{medhat2017service} which chains different VNFs together. Some VNFs may deliver the packets they have processed to multiple possible follow-up VNFs, which depend on the content of the processed packets. As shown in figure 1, if the packet which goes through the web firewall contain some malicious patterns, such as \textit{"1 = 1"}, which may be used to conduct SQL injection attack\cite{halfond2006classification}, such packets will be delivered to Intrusion Detection System (IDS) by invoking $ECALL\_IDS$. In this case, the enclave interface invocation sequence is $ECALL\_FW$, $ECALL\_IDS$. Otherwise, the packet will be delivered from the web firewall to NAT directly. In this case, the enclave interface invocation sequence is $ECALL\_FW$, $ECALL\_NAT$. By observing which interface invocation sequence occurs during processing a packet, the attacker can infer whether this packet contains content "1=1".

\begin{figure}[!h]
\begin{lstlisting}[ language=C++] 
int ECALL_WF( PKT pkt ) { 
    string url = pkt.get_url();    
    if ( url.find( "1=1" ) ) {
        ECALL_IDS( pkt );
    }
    else {
        ECALL_NAT( pkt );
    }
    return 0;
} 
\end{lstlisting} 
\caption{Example Function with input-dependent control transfer}
\end{figure}
 
\subsubsection{interface parameters}
The data transferred from memory to enclave is always encrypted. However, the size of the encrypted data is not always concealed\cite{han2017sgx, kim2017enhancing}. The length of the ciphertext is always proportional to the corresponding plaintext, which can be observed when the data delivered as parameters of ECALL/OCALL interface. Thus the attacker can profile different input data by the length of the encrypted input data. Specifically, there are two kinds of side channel information caused by the enclave interface parameters. The first type is the static parameter pattern of ECALLs/OCALLs, which is the size of each parameter of the ECALL/OCALL. The second type is dynamic parameter patterns, which is the change of the parameter size of continuous ECALLs/OCALLs.

For example, different images or network packets have different sizes, which can be exploited by attacks to profile images or packets. In addition, some applications will change the input data, which will cause the change of the data size. For example, the data compression application will compress the data and change the data size. As shown in the Figure 2, the compression function will compress the input data \textit{data} and deliver it to the codes outside the enclave such as network I/O. By observing the size of \textit{data} and \textit{cprs\_data}, an attacker can infer whether the input data is text or a picture because the gzip algorithm has the compression efficiency which is larger than 30\% for texts, but less than 5\% for .png or .jpg pictures based on our tests. Furthermore, different data always causes different size changes. If an attacker can know the compression ratio of different input data, the attacker can track the input data by observing the compression ratio of the enclave input data.

\begin{figure}[!h]
\begin{lstlisting}[ language=C++] 
int ECALL_CPRS( string data ) {
    string cprs_data = compress( data )
    OCALL_DLV( cprs_data )
    return 0;
}
\end{lstlisting} 
\caption{Example Function of Gzip Compression}
\end{figure}

\subsubsection{interface invocation delay}  
Different enclave input data will always trigger different program control flows, which contain different instructions. Thus it will take different time to finish tasks of processing different enclave input data. By calculating the time between the invocations of ECALLs which used to start the task and subsequent OCALLs, an attacker can know the time which is used to process the input data. This time can be used to profile different enclave input data.

For example, the text segments with different contents usually need different time to finish the regular expression matching tasks. The attack can use different regular expression matching time to profile different text segments.

\subsection{Interface Invocation Patterns Collection} 
To conduct interface-based side channel attack, an attack needs to collect the interface-based side channel information mentioned above. In this section, we illustrate how to collect three types of interface-based side channel information.

\subsubsection{collecting interface invocation sequences}
To collect the interface invocation sequences, we need to intercept all of the ECALLs and OCALLs in the OS, record the invocation information and execute the ECALLs and OCALLs subsequently.

For ECALLs, Intel SGX SDK uses \textit{edger8r} tool to convert the user-defined ECALL interfaces into the unified interface \textit{sgx\_ecall()} during building. \textit{sgx\_ecall()} is defined in the shared library \textit{libsgx\_urts.so} outside the enclave, with a parameter distinguishing different ECALLs. Thus we use run-time interpositioning\cite{bryant2003computer} to intercept the calls to \textit{sgx\_ecall()} from untrusted codes. We set the \textit{LD\_PRELOAD} environment variable to our own shared library which has the same name as \textit{libsgx\_urts.so} and contain a function with the same function declaration as the real \textit{sgx\_ecall()}. In our fake \textit{sgx\_ecall()}, we record the parameter which distinguishes different enclaves and ECALLs, record the CPU clock cycle and then call the real \textit{sgx\_ecall()} to enter the enclave.

As for OCALLs, we need to use a different way to intercept them, which we call it ocall\_table hijack. The OCALLs are invoked inside of the enclave. Thus every time an ECALL is invoked, the \textit{sgx\_ecall()} will use the parameter \textit{ocall\_table} to indicate the addresses of the OCALLs which can be invoked by codes inside of the enclave. As shown in figure 3, the \textit{ocall\_table} contains 
\textit{n} OCALL addresses. We construct a data structure \textit{hjk\_ocall\_table} in the memory whose layout is the same as \textit{ocall\_table}. Then use\textit{hjk\_ocall\_table} to replace the \textit{ocall\_table} in \textit{sgx\_ecall()}. The addresses in \textit{hjk\_ocall\_table} direct to corresponding trampoline codes, which record the index of corresponding OCALL in the register and direct to our \textit{rcd\_fun()}. The \textit{rcd\_fun()} will record the CPU clock cycle, record the OCALL index in the disk, calculate the address of the real OCALL address and jump to it. 

\begin{figure}
\includegraphics[width=2.0in]{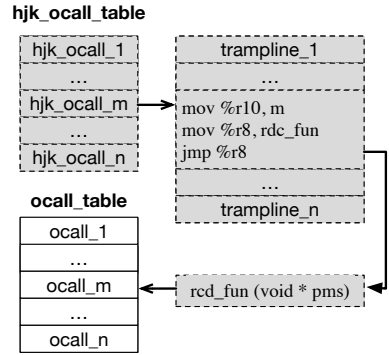}
\caption{Ocall\_Table Hijack}
\end{figure}

\subsubsection{collecting parameter size}
For every ECALL/OCALL, they all use a point parameter \textit{pms} to point to a struct \textit{ms} which contains their parameters. We can get \textit{pms} in the register when we intercept each ECALL/OCALL. But we don't know the data structure of each interface's \textit{ms}. However, the SGX SDK will add parameter unmarshaling codes before the execution of ECALL/OCALL. Thus we can analyze the source codes or disassemble the binary codes of unmarshaling logic to gain the detail of the \textit{ms} and get the parameter size.

\subsubsection{collecting interface invocation delay}
We use the RDTSC instruction to record the number of clock cycles before the invocation of each ECALL in our run-time interpositioning program. We also record the clock cycles before subsequent invocation of OCALL in our \textit{recording\_function}. The different between these two number can present the invocation delay of each ECALL. 

However, the system events will generate noises for delay measurement. To reduce the noise, we isolate one cpu core for the application and minimize the cpu frequency. Plus, we schedule the the eliminable system interrupt to other cpu cores. 

\section{Atack on NFV Platform}
In this section, we conduct an interface-based side channel attack in an SGX-assisted NFV platform which is used to process the web traffic in an enterprise. In this attack, the attacker can infer whether the tracked website pages have been visited by enterprise hosts and identify the packets which belong to the tracked websites, which breaks the privacy-preserving guarantee of such platforms. For simplicity, we mainly focus on the encrypted HTTP web traffic in the following. However, the attack for HTTPS web traffic is similar, because HTTPS also use TLS as secure transmission protocol as the enterprise gateway does under our assumption.  

\subsection{SGX-assisted NFV Platform}
Network function virtualization\cite{chiosi2012network}(NFV) is the technology that runs the software middleboxes in virtual machines on the standard servers. Enterprises favor NFV to build their network for its benefits such as low equipment cost and energy consumption, optimizing network configuration real time and supporting multi-tenancy, etc.

The public cloud is an alternative place to deploy NFV platform in addition to enterprises' local data centers. Outsourced NFV platform has been embraced by enterprises for its enormous benefits such as reducing the management and maintenance difficulties, resource elasticity, and democratizing access to middlebox services by pay-per-use model.

However, the resources in the cloud are not under the enterprises' control, there are risks of exposing the privacy of users' traffic which goes through the VNFs in the cloud. Malicious staff in the cloud can steal the privacy of traffic easily. Research has also shown various attacks that can steal private data from cloud environments\cite{bugiel2011amazonia,ristenpart2009hey,zhang2011homealone}. Confronted with the untrustworthiness of cloud platforms, SGX-assisted privacy preserving NFV platforms\cite{kim2015first,trach2018shieldbox,han2017sgx,shih2016s,kohler2000click,duan2017lightbox} have been proposed to protect the privacy of traffic which goes through the outsourced NFV platforms. 

The structure of the SGX-assisted NFV platform is shown in Figure 4, enterprises can ensure the specified virtual network function (VNF) enclave has been initialized correctly in the cloud through SGX remote attestation mechanism\cite{anati2013innovative}. After remote attestation, an enterprise gateway can establish a trusted encrypted channel between the enterprise gateway and VNF enclave through secure transmission protocol, such as TLS. When the network traffic reaches to the enterprise gateway, it will be encrypted and redirected to the privacy-preserving NFV platform in the cloud by the enterprise gateway. The network I/O of untrusted OS in the cloud will receive and deliver all the received encrypted packets to SGX-assisted VNF. After entering the enclave of SGX-assisted VNF, the encrypted packets will be decrypted. Then the packets can be processed by VNF as plaintext in the enclave. Finally, the packets will be encrypted again in the enclave and sent back to the enterprise gateway through the network I/O of the untrusted system. Because the encrypted packets are only decrypted in the enclave which is trusted, the privacy of the network traffic content can be ensured. But current SGX-assisted NFV platforms fail to obscure interface invocation patterns of the enclave, which can be exploited to conduct the interface-based side channel attack.

\begin{figure}
\includegraphics[width=2.5in]{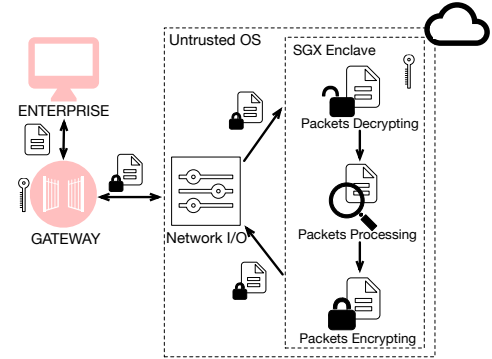}
\caption{SGX-assisted privacy preserving NFV platform}
\end{figure}

\subsection{Key Oberservations}
There are some key observations which can be used to conduct our interface-based side channel attack on SGX-assisted NFV platform. 

First, The content of a particular web page will not change in a short time. Therefore, the content and size of request packets of objects in the particular web site will not change in a short time. Second, different web pages have different content such as HTML layout, images, flashes, etc. Before the requested objects of web pages are transmitted, most of them are split into several TCP segments with different sizes. Each web page will have some objects which always have the stable split segments. The size and content of these segment packets are determined. Thus, these constant packets whose contents are constant can be used to profile each web page. Third, when one VNF processes the same packet multiple times, they will expose the same interface-based side channel information. So we can infer whether a tracked website page occurs by identifying whether the constant packets which belong to the website page has occurred. And each packet can be distinguished by the interface-based side channel information exposed by the SGX-assisted NFV platform. 

Based on the key observations and attack overview mentioned above, we show the details of three phases of the interface-based side channel attack on SGX-assisted NFV platforms in the next three parts.

\begin{table*}
	\caption{Exploitable Features Exposed by Typical VNFs}
	\begin{tabular}{ccl}
		\toprule
		 VNF & Interface-based Features & Meaning \\
		\midrule
		IP Firewall & Parameter Size, Invocation Delay, Invocation Sequences & Packet Size, Processing Delay, Packet Delivery Path \\
		NAT & Parameter Size, Invocation Delay & Packet Size, Processing Delay \\
		IDS/IPS & Parameter Size, Invocation Delay, Invocation Sequences & Packet Size, Processing Delay, Disk Read/Write \\
		WAN Optimization & Parameter Size, Invocation Delay, & Packet Size Change, Processing Delay \\
		Parental Filter & Parameter Size, Invocation Delay, Invocation Sequences & Packet Size, Processing Delay, Packet Delivery Path \\
		Load Balance & Parameter Size, Invocation Delay & Packet Size, Processing Delay \\
		Proxy/Cache & Parameter Size, Invocation Delay & Packet Size, Processing Delay \\
		\bottomrule
	\end{tabular}	
\end{table*}

\subsection{Offline Pattern Collection}

In the pattern collection phrase, we will build profiling set \textit{P} for each tracked web pages in dataset \textit{D}. Each profiling set contains \textit{n} constant packets $\{p_0,p_1,...,p_n\}$ which can be used to identify the tracked pages. \textit{n} varies as web page changes. These packets include all the request packets and some object response packets whose content is decisive. We can figure out these packets by visiting each tracked web pages n times and record the packets which appears every time. Each packet $p_i$ is represented as vector composed by a set of interface-based side channel features $\{f_{i0},f_{i1},...,f_{im}\}$, \textit{m} varies when the topology of the SGX-assisted NFV platform changes. We build \textit{k} profiling set \textit{P} for each website page, which contains the packets sequence information of each website page.

To build the side channel feature vector for each packet, we need to know what interface-based side channel information will expose when a packet goes through each kind of VNF.

The observable interface invocation sequences on the SGX-assisted NFV platforms include the system calls of file operations triggered by different packets and the ECALL/OCALL invocation sequences during the delivering of different packets. When VNFs process the packets whose contents are suspicious or are interested in by network managers, the log function will be triggered. Such log functions will call the file system I/O interface to interact with the file system. Thus, different packets will cause different disk read or write operations. From the invocation of specified system calls, the attacker can know that the packet contains the content which can trigger the log function. In addition, when different packets go through privacy preserving NFV platform, they may go through different forwarding path. For example, when the packet which contains the suspicious content or IP addresses goes through the WAF in Figure 7, it will be forwarded to IDS to perform deep packet inspection. Otherwise, it will be forwarded to NAT directly. From the different packets delivery path, the attacker can distinguish different packet easily.

The parameter size of interfaces on the SGX-assisted NFV platforms is the size of encrypted packets which need to be copied into the enclave. Although the encryption can conceal the content of packets, the AES256 encryption algorithm used in TLS during secure transmission cannot conceal the size of the packets. The size of each encrypted packets is decisive and unique. Furthermore, when a packet goes through some VNFs which will change the content of the packets, the packets size will change. For example, the WAN OPT is used to compress the content of the packet to reduce the packet transmission delay on the internet. Different contents will cause different size changes. 

Different VNFs process packets according to their rules, so different packet will cause different processing logic and generate different processing delay. And every kind of VNF will cause this kind of side channel information. We represent the ECALL invocation delay of each packet in the profiling set as a range whose lower limit and upper limit are the shortest and the longest ECALL invocation time interval during multiple processing. We consider a new packet's interface invocation delay feature matches with the interface invocation delay feature of a packet in the profiling set only if the new packet's interface invocation delay feature falls into the interface invocation delay range of the packet in the profiling set.

In summary, different VNF will expose different kinds of side channel information when processing packets. Table 1 shows the side channel information exposed by typical VNFs deployed in enterprise networks. Almost all kinds of VNFs expose interface invocation delay information. Firewall and Parental Filter will expose packet delivery path information through interface invocation sequences because these two kinds of VNFs are always connected with IDS/IPS. Serving as a monitor, IDS/IPS, and Parental Filter will expose disk read or write information when they log the needed information through interface invocation sequences. WAN OPT will expose the packet size change information through size change of interface parameters because it will change the contents of packets. 

In our attack, we use the interface-based features of each VNF shown in Table 1 to construct the feature vector of each packet. It includes the interface invocation delay and parameter size of interfaces of WAF, NAT, IDS and WAN OPT, the interface invocation sequences of the whole platform, the write operation of IDS and the packets size change of WAN OPT.

\subsection{Offline Algorithm Design}
In the algorithm design phrase, we design two algorithms to infer whether the tracked web site pages have been processed by SGX-assisted NFV platform and identify the packets which belong to the tracked web site pages. Then we train the algorithm using the profiling set \textit{P}.

\subsubsection{Page Recognition Algorithm}
After gaining the feature vector $p^{'}_{i}$ for each new packet, we need to figure out which tracked web page this packet belongs to and store the result. To perform this task, we create a matching progress indicator \textit{I} for each tracked web page. The matching indicator $I_i$ for each tracked web page $W_i$ indicates the appearing number of each packet in its profiling set $P_i$. When feature vector $p^{'}_{i}$ of a packet matches with a feature vector $p_i$ in a profiling set $P_i$, we will add the account of $p_i$'s appearing times by 1 and calculate the ratio of appeared packet $R_{appeared}$ in profiling set $P_i$ using formula 1. $N_{appeared}$ represents the number of packets whose appearing number in $I_i$ is not zero. $T$ represents the total number of packets in the profiling set $P_i$. When the $R_{appeared}$ reaches 100\%, we can infer that the corresponding tracked web page has been visited. In addition, we will store the information $m_i$ about $p^{'}_{i}$ with the index of $p_i$ in the information buffer \textit{B}. This information includes the feature vector $p_i$, its arriving time and arriving sequence number. This information will be used in the packet recognition. We set a 30s timer for each $m_i$. The $m_i$ will be cleared when its timer is used up, which is used to save storage resources. 

\begin{equation}
	R_{appeared} = \frac{N_{appeared}}{T} 
\end{equation}

For example, as Figure 5 shows, after training phrase, the profiling set $P_i$ which contains four packets feature vector for tracked web page $W_i$. When the matching progress indicator $I_i$ is initialized, the appearing time of all the packets are zero. Then we will compare feature vector of the packet in the web page traffic with feature vector in $P_i$ one by one. After processing \textit{t} packets, the state of \textit{I} has changed. And it indicate that packet $p_1$, $p_2$, $p_3$, $p_4$ has appeared 2, 4, 3, 0 times respectively. And the $R_{appeared}$ is equal to 75\%. After matching the feature vector of \textit{t+1$_{st}$} packet with all the feature vector in every profiling set, we find $p^{'}_{t+1}$ is equal to $p_4$ in \textit{$P_i$}. Then the appearing times of $p_4$ in \textit{$P_i$} will change to 1. And the $R_{appeared}$ will change to 100\%. This means the web page presented by \textit{S} has been visited.

\begin{figure}
\includegraphics[width=2.5in]{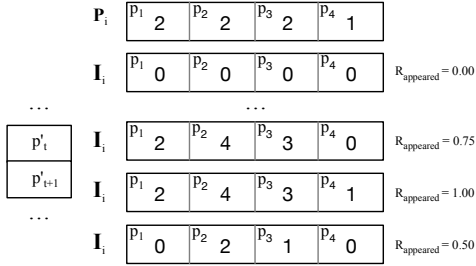}
\caption{Recognition Algorithm Example}
\end{figure}

We infer whether a tracked web page has been visited through the variation of the value of the $R_{appeared}$. However, when the tracked web page has been identified the $R_{appeared}$ will continuous remain 100\%. Thus we need to clear the states produced by packets of the identified web page. The traffic may contain the packets which are generated by visits the same web page multiple times and the same web page may have multiple packets whose features vector are the same. Thus we cannot clear the value of \textit{I} naively by setting the appearing times of all packets to 0 or reduce 1 from the appearing times of each packet. We need to reduce the times which they appear during visiting web page once from appearing times of each packet. We can get the appearing times of each packet in profiling set through recording the appearing times of each packet when building the profiling set. As shown in Figure 5, if the appearing times of each packet in \textit{P} is 2, 2, 2, 1 respectively, then after clearing, the \textit{I} will become 0, 2, 1, 0. $R_{appeared}$ changes to from 100\% to 50\% correspondingly.

\subsubsection{Packet Recognition Algorithm}
After the page recognition, we can know whether the network traffic of a tracked website page has been processed by an SGX-assisted NFV platform. However, we cannot figure out whether a packet belongs to the tracked web site page. For example, As shown in figure 5, when the $R_{appeared}$ is equal to 1, there may be 2 $p_1$, 4 $p_2$ and 3 $p_3$ in the information buffer\textit{B}, we need to figure out the packets which truly belong to the tracked website page from all these packets. 

 The packets which belong to the same web page always have the stable relative positions in the web page traffic. For example, the request packets usually occur before corresponding response packets. And most web pages finish their rendering in \textit{t} seconds which is the max time interval between any 2 packets in each tracked web site's profiling set. Thus the arriving time intervals between the packets in the same web page traffic will less than \textit{t} seconds. We use the packets sequence properties of each tracked website and arriving time interval of the packets in each tracked website to identify the real packets.

\begin{figure}
\includegraphics[width=2.0in]{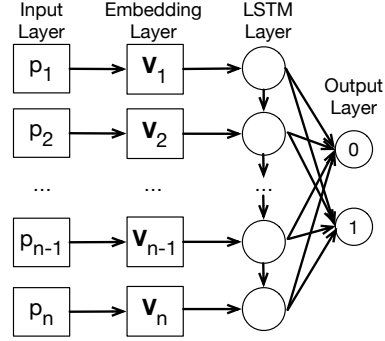}
\caption{Long Short Memory Network}
\end{figure}
 
 The Long Short Memory Network (LSTM) is a classifier which will consider the context when it processes each input data. Thus we can use it to predict whether a packets sequence is legal for a tracked network web page. As shown in Figure 6, we can use the packets sequence as the input of the binary classification LSTM which can predict whether the input packets sequence is legal of a tracked web site page. The output layer will show the probability of whether the input packets sequence is legal. We first use the packet sequence in \textit{k} profiling sets of each tracked website page to train their own LSTMs. Then we use all possible packets sequences in the matching indicator $I_i$ whose $R_{appeared}$ is equal to 1 as input to predict their legality. We add the legal packet sequence into legal sequence set \textit{L} for each tracked web site page. 
 
 Before we use each tracked website page to predict the legality of each packets sequences, we use the time interval \textit{t} to filter some illegal packets. Firstly, we compare the arriving time of each packet in the matching indicator $I_i$. If the arriving time intervals between a packet and all other packets are all large than \textit{t}, this packet must not belong to the tracked web site page. Secondly, if the arriving time interval between any 2 packets in the sequence is larger than \textit{t}, this sequence is illegal. Finally, the packets contained in the packets sequences left in \textit{L} may belong to the tracked web site pages.

\subsection{Online Data Inferring}
In the online data inferring phrase, we slow down the delivery rate between the network I/O and VNF enclave to make the VNF process the packet in the web traffic one by one. After observing the interface-based side channel information exposed by each VNF when packets go through the whole NFV service function chain. We can obtain the feature factor of each packets $p_{i}^{'} = \{f_{i0}^{'},f_{i1}^{'},...,f_{im}^{'}\}$. And when we collect the side channel information exposed by the VNFs, we need to make sure the VNF starts to process the next packet until the previous packets have been delivered. If multiple packets are processed in VNF simultaneously, we will not match the input packet with output packet because the content of a packet will change after being encrypted again. Then we use web page recognition algorithm and packet recognition algorithm to infer whether tracked website page has been processed by SGX-assisted NFV platform and identify the packets which belong to the tracked web pages.

\section{Countermeasures}
Through the attack we conduct above, an attacker can know that whether the tracked web page has been visited by the hosts in the enterprise. By tracking the privacy-related website pages, such as the particular medical web page or shopping web page, an attacker can infer the health conditions and shopping habits of employees in enterprises. The privacy of enterprise employees leaks seriously, which break the guarantee of SGX. Confronted with three kinds of interface-based side channels in SGX-assisted applications, we propose following countermeasures to reduce the ability of these interface-based side channels and improve the security of SGX-assisted applications. The countermeasures include fixing the size of input data and batch operation mechanism.

\subsection{Fixing Size of Input Data}
Different enclave input data may have different sizes, which can be observed as parameter size and be used to profile different enclave input data. For example, different web page traffic contains packets of different sizes. Thus, the size of the constant packet in web page traffic can be used to profiling tracked web page. To reduce the side channel information caused by the interface parameter sizes, we need to fix the size of the interface parameter size which is various as input data change. Before the sensitive data is encrypted and sent to an untrusted party, we pad the data to a fixed length, which makes the ciphertexts have the same length. There are several strategies to pad the data. For example, we can pad the input data length to the max length of all the input data or to \textit{x} times the fixed length. \textit{n} is the least positive natural number which satisfies $n*x > L$. \textit{L} is the real length of the ciphertexts. Different strategies have different cost and security effectiveness. Padding the data length to the max length of all of the input data will waste large amounts of resources, such as network bandwidth, but leak the least side channel information, padding the size to \textit{x} times the fixed length depends on the real length of input data, which will waste little resources but reveal more side channel information about the input data size.

Through fixing the length of enclave input data before sending it to the untrusted party and sending them out of SGX enclave, the interface parameter size information observed by an attacker will be reduced. Such information will make little sense when being used as features to identify particular enclave input data. Therefore, side channel information of parameter size can be reduced. 

\subsection{Batch Operation}
Confronted with side channel information of interface invocation sequences and interface invocation delay, we propose the batch operation mechanism to reduce such information, it includes batch data delivery and batch interface invocation. 

To gain the interface invocation sequences and interface invocation delay, an attack needs to process the enclave input data one by one. If we deliver the input data in batch, these two kinds of interface-based side channel information cannot be collected precisely. For example, we assume the batch delivery threshold is \textit{n}. With batch delivery mechanism, we can ensure that all the encrypted data is delivered into enclave in a batch of \textit{n} data. Similarly, only after all the \textit{n} data being processed, can they be sent out of the enclave together. By this way, all the data is processed in the enclave simultaneously. Thus the attacker cannot know the exact processing delay of each input data. In addition, the interface invocation side channel information will also be reduced under the batch delivery mechanism. The attacker cannot map each output data in the batch with each input data in the batch after encrypting the packets in enclave again. Thus the attacker cannot know the interface invocation sequences of each input data. In terms of implementation, we cannot only implement the batch delivery mechanism outside the enclave, because the system is untrusted. We need to ensure that the packets are delivered in batch in the enclave. The batch operation mechanism in the enclave will verify the input data to ensure that the codes in enclave will not execute until receiving a batch of packets. 


\section{Evaluation}
In this section, we show the effectiveness of the interface-based side channel attack on SGX-assisted NFV platform which is used to process the web traffic and the effectiveness of our countermeasures. We first describe the data set which will be used to perform the attack and evaluate the effectiveness of countermeasures. Then we describe our experiment setup which is used to evaluate the effectiveness of the interface-based side channel attack and the effectiveness of the countermeasures. Finally, we show the experiment result of our attacks under different conditions and the effectiveness of countermeasures. Through the following experiments, we show that our attack can identify tracked website page with the accuracy of 87.6\% and recall of 76.6\% and identify the packets which belong to the tracked web pages with the accuracy of 67.9\% and recall of 71.1\%. And our countermeasures will make the accuracy of our attack reduce to 0.

\subsection{Data Collection}
To illustrate the effectiveness of the interface-based side channel attacks and countermeasures, we need to build two data sets. One is the training data set, and the other is the test data set. The training data set is the packet set which contains the traffic of each tracked web pages. It is used to build the profiling set for each tracked web page. The test data set contain the packets of both tracked web pages and other untracked packets. It is used to evaluate the effectiveness of attack and countermeasures. We collect 2500 URLs from 200 famous HTTP websites from Alexa websites rank. These websites are distributed in 10 different countries located in 5 different continents, such as the US, England, China, Australia, etc. 

To build the training data set, we choose \textit{m} URLs from 2500 URLs and visit each of tracked URLs 20 times. We use tcpdump to collect the TCP packets whose size is larger than 100 bytes during visiting each URL. Those packets whose sizes are smaller than 100 bytes are used to establish the TCP connection. They will not make sense because they will not contain any meaningful contents. \textit{m} will change according to our different experiment intentions. After building the training dataset, we can process the packets in training data set by the service function chain we built. During each packet processed by VNFs, we can gain the side channel information and built profiling set for each tracked web page.

To build the test data set, we need to collect the traffic which is generated by visiting multiple web pages simultaneously. Thus, we choose \textit{n} URLs from 2500 URLs and use tcpdump to capture the traffic when visiting each of them twice simultaneously. The \textit{n} URLs in the test dataset include the \textit{m} URLs in the training dataset. However, we cannot visit \textit{2n} web pages simultaneously because of our limited memory. We visit 500 URLs simultaneously one-time using Chrome browser and repeat this procedure several times to finish constructing traffic contain \textit{2n} web pages. 
 
\begin{figure}
\includegraphics[width=2.5in]{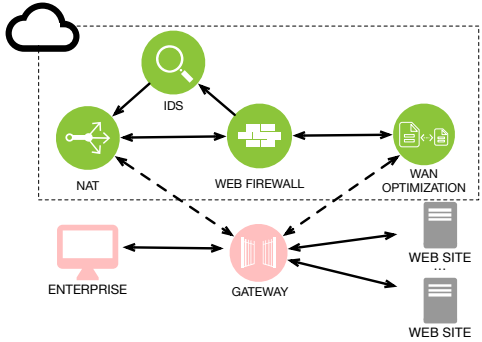}
\caption{System architecture of privacy preserving NFV platform used to process Web traffic}
\end{figure}

\subsection{Experimental Setup}
We use the topology shown in Figure 7 as the service function chain  containing typical VNFs\cite{kumar2015service, sherry2012making} which are deployed by enterprises to process the web traffic. When a host visits a website, the request packets will be redirected to the NFV platform in the cloud and go through Network Address Translator (NAT), Web Application Firewall (WAF) and Wan Optimization (WAN OPT) sequentially. NAT is responsible for the translation between the LAN IP addresses and WAN IP addresses. Web Firewall is used to perform access control, Cross Cite Script（XSS）Detection, etc. To decrease the volume of packets transmitted on the Internet and reduce the packets transmission delay, WAN OPT is used to compress and decompress the Web traffic packets. After the response packets arriving at enterprise gateway, they will also be redirected to NFV platform on the cloud. The response packets may contain the malicious executable codes. Thus the suspicious packets will be forwarded to the Intrusion Detection System (ISD) after processed by WAF. IDS will perform deep packet inspection to confirm whether the packet is malicious. The network setup not only can process HTTP web traffic, when the NVF support TLS\cite{paladi2017safeguarding,han2017sgx}, it can also process HTTPS traffic. As most privacy preserving NFV platforms\cite{kim2015first,trach2018shieldbox,han2017sgx}, we use TLS as secure transmission protocol between enterprise gateway and privacy-preserving NFV platform.

We implement four SGX-assisted VNFs and the enterprise gateway in Figure 7, including a binary trie based\cite{ruiz2001survey} NAT, a pattern matching based IDS, and WAF, and a gzip-based WAN OPT. The IDS uses 1k-5k rules. The WAF uses 1k rules used for web attack detection. The enterprise gateway and VNFs use AES 256 to encrypt the packets when transmitted between various VNFs outside the enclave. 

All experiments are conducted on the desktop with 4 core 8 threads Intel Core i7-6700 CPU (3.4GHz) and 8GB memory, running Ubuntu 16.04 TLS and SGX SDK v2.0. We use gcc 5.4.0 to compile the SGX-assisted VNFs. The enclave memory is set to the maximum 128MB. Each NFV in our experiment runs as a separate progress.



\begin{figure*}
\centering
\begin{minipage}[t]{0.33\textwidth}
\centering
\includegraphics[width=6cm]{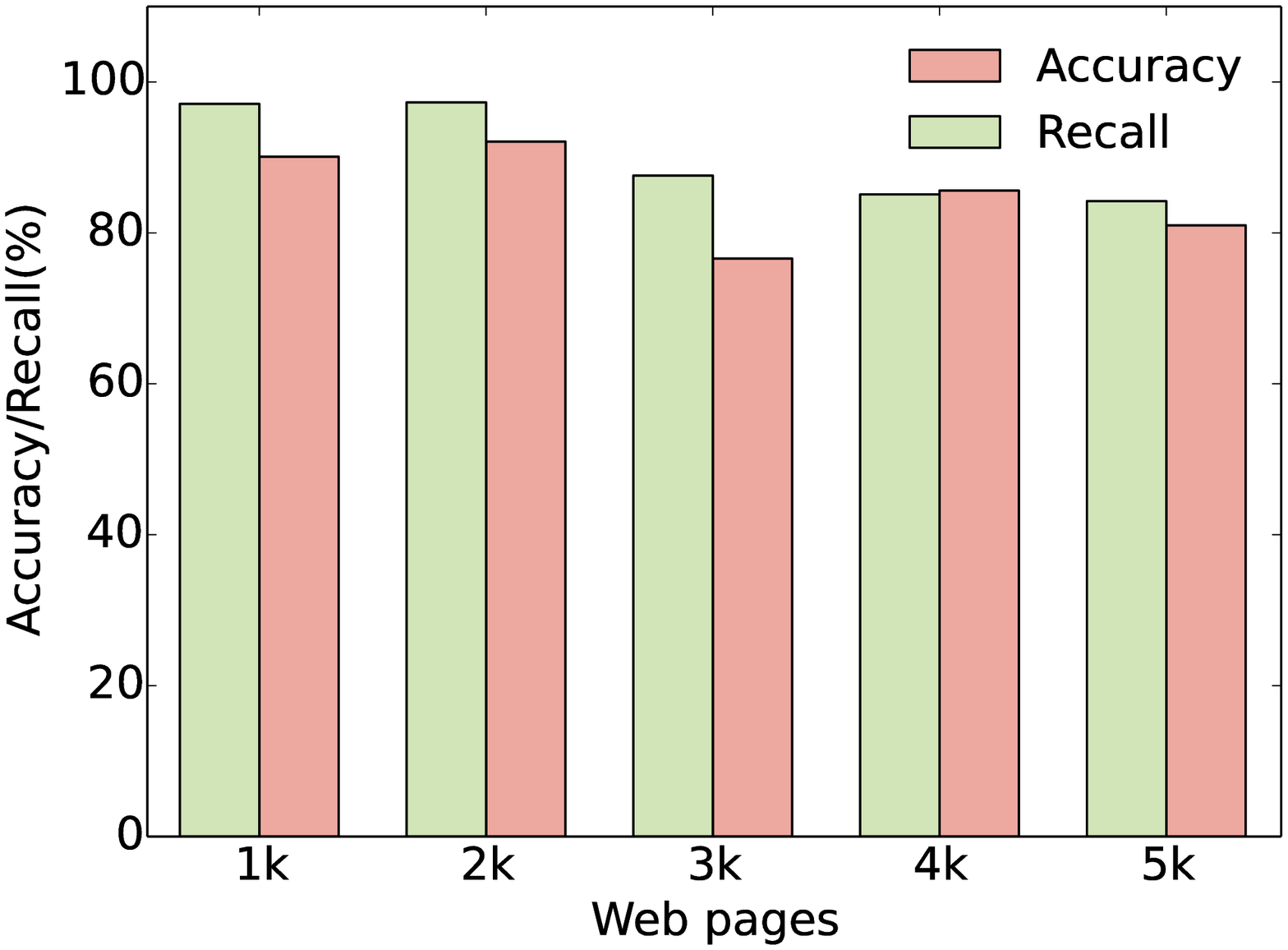} 
\caption{Web Pages Recognition Performance(50\% of tracked web pages)}
\end{minipage}
\begin{minipage}[t]{0.33\textwidth}
\centering
\includegraphics[width=6cm]{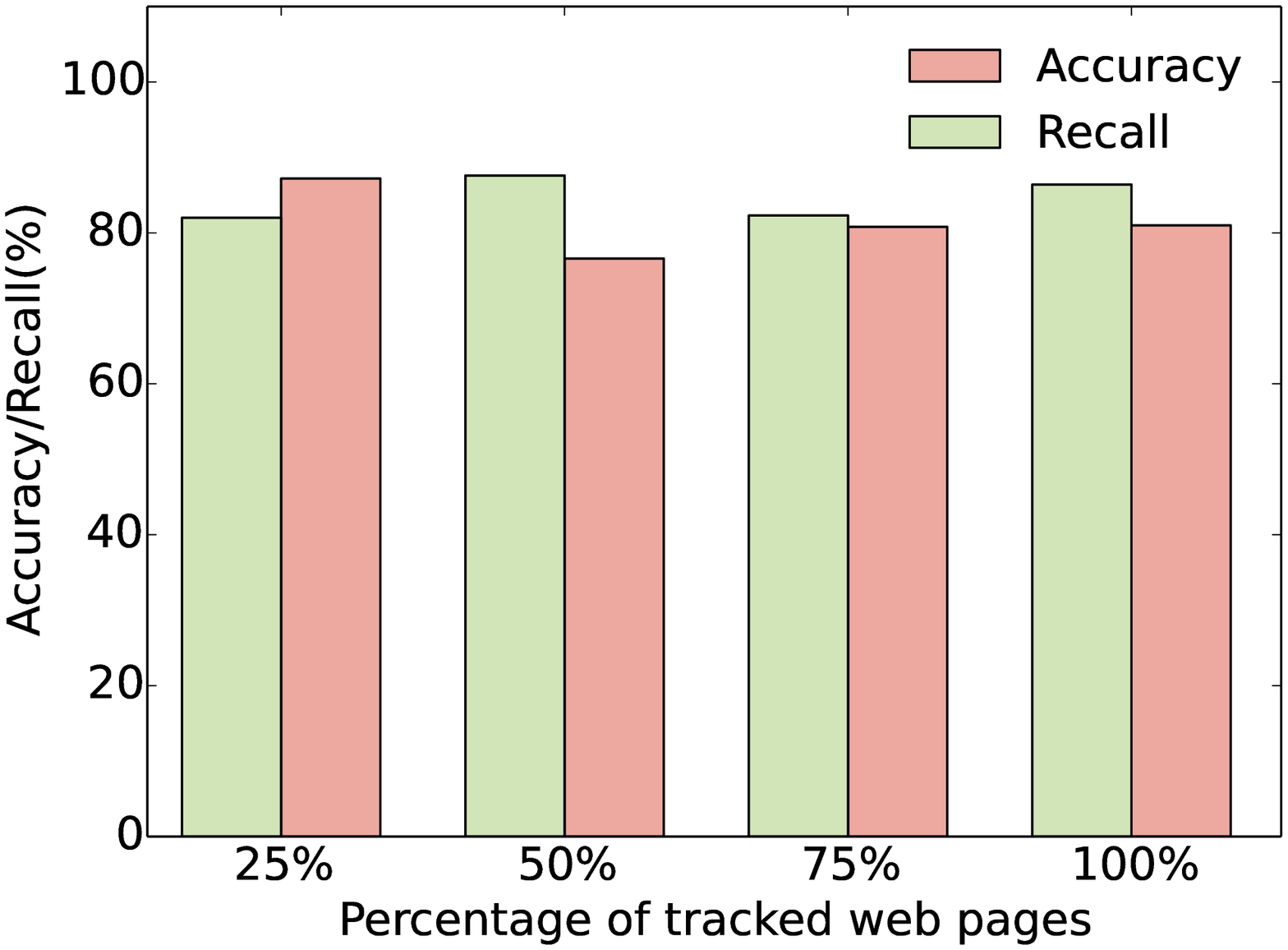}
\caption{Web Pages Recognition Performance(3k web pages)}
\end{minipage}
\begin{minipage}[t]{0.33\textwidth}
\centering
\includegraphics[width=6cm]{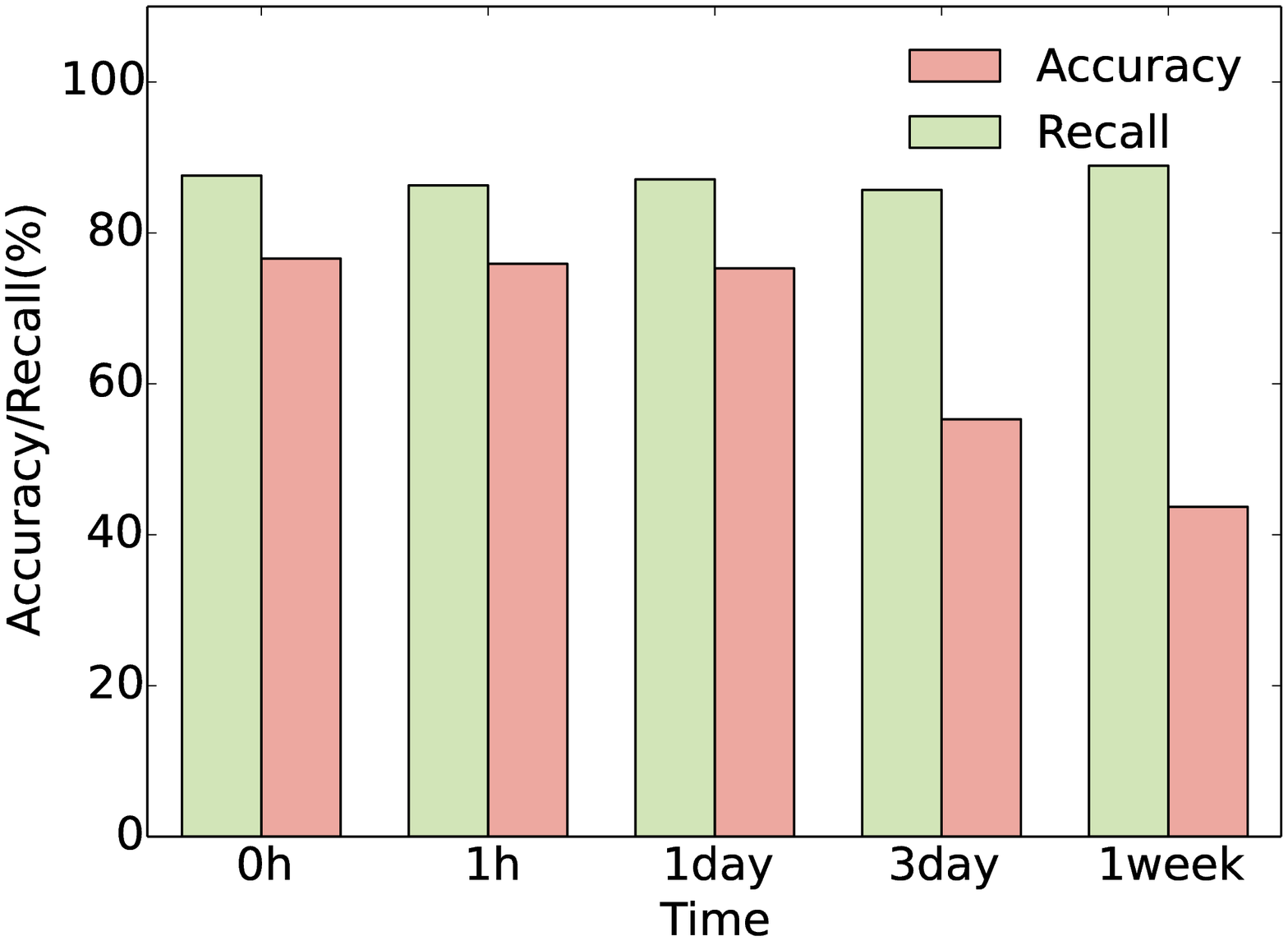}
\caption{Web Pages Recognition Performance(3k web pages with 50\% tracked web pages)}
\end{minipage}
\end{figure*}

\begin{figure*}
\centering
\begin{minipage}[t]{0.33\textwidth}
\centering
\includegraphics[width=6cm]{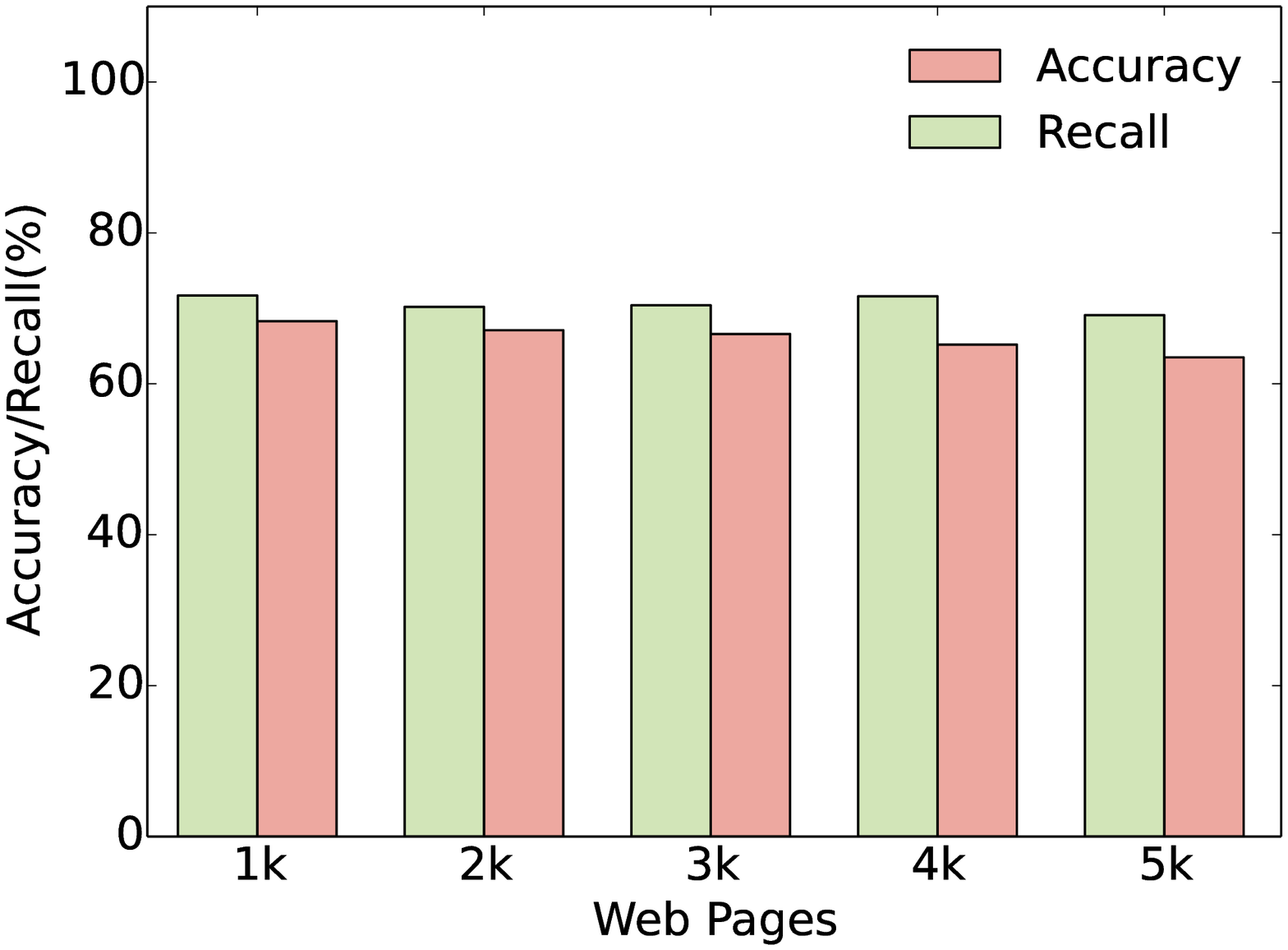}
\caption{Packets Rocognition Performance(50\% of tracked web pages)}
\end{minipage}
\begin{minipage}[t]{0.33\textwidth}
\centering
\includegraphics[width=6cm]{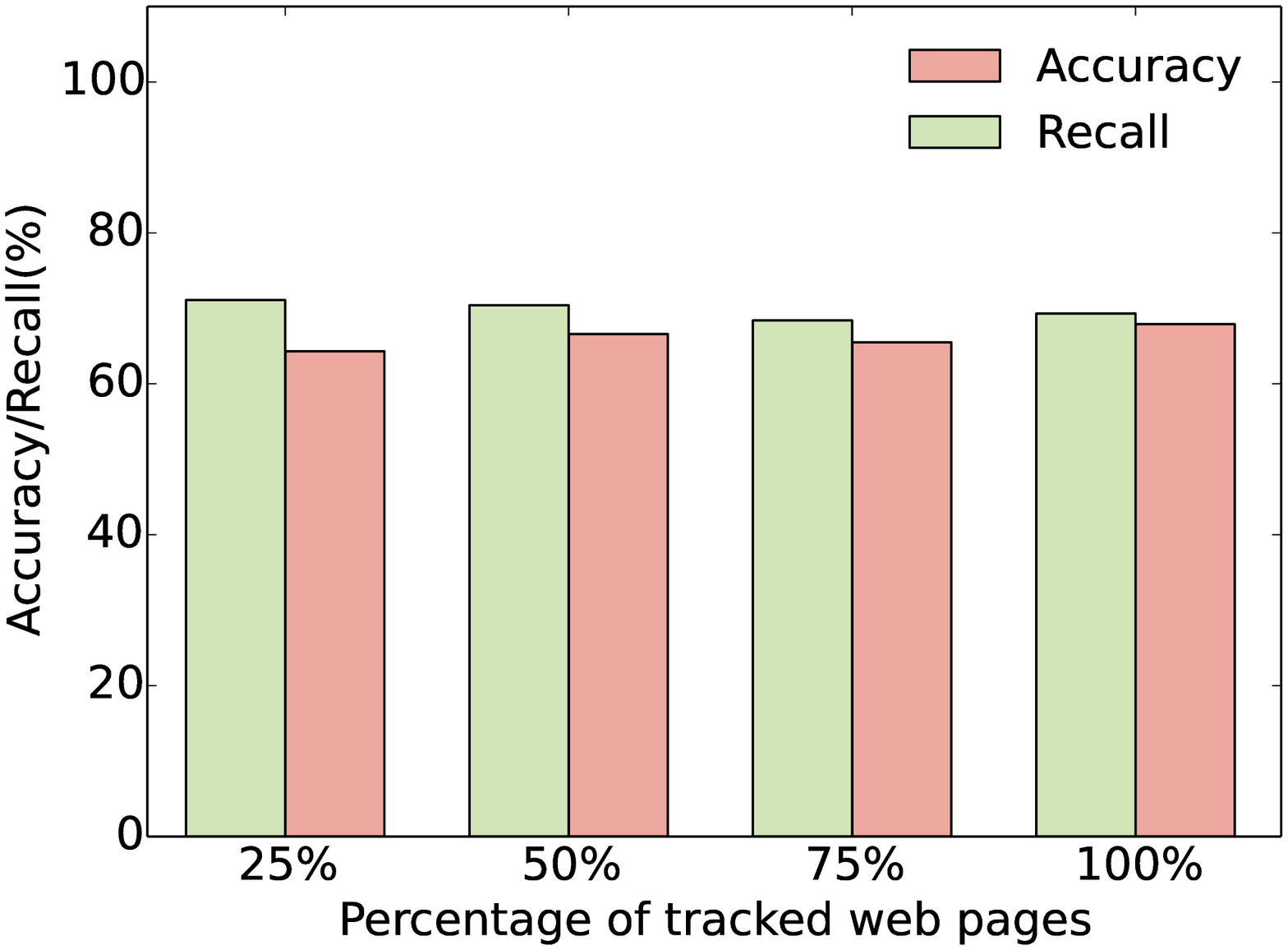}
\caption{Packets Rocognition Performance(3k web pages)}
\end{minipage}
\begin{minipage}[t]{0.33\textwidth} 
\centering
\includegraphics[width=6cm]{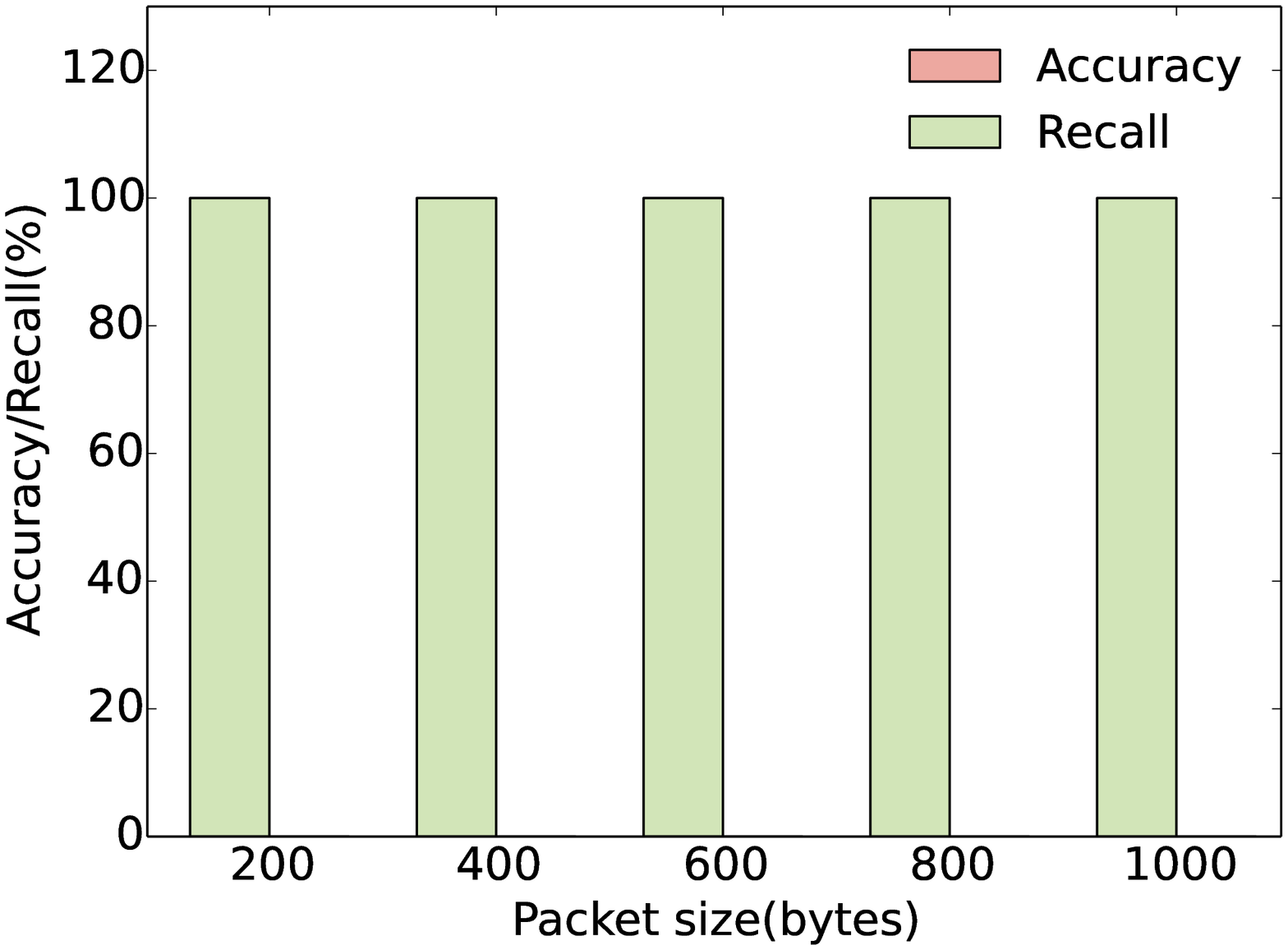}
\caption{Countermeasures Effectiveness(batch processing and 600 bytes packets size)}
\end{minipage}
\end{figure*}

\begin{figure*}
\centering
\begin{minipage}[t]{0.33\textwidth}
\centering
\includegraphics[width=6cm]{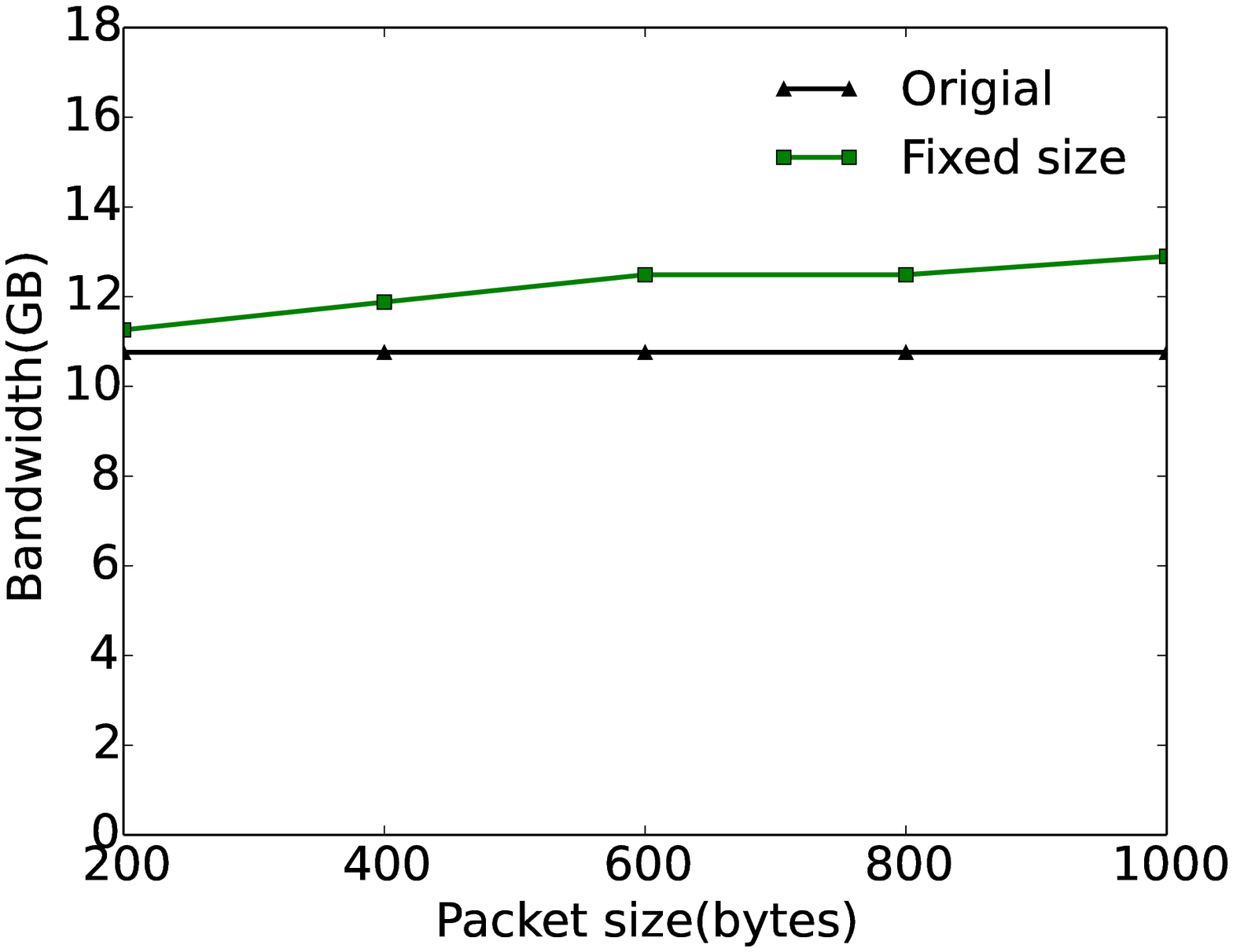}
\caption{Fixed Size Overhead}
\end{minipage}
\begin{minipage}[t]{0.33\textwidth}
\centering
\includegraphics[width=6cm]{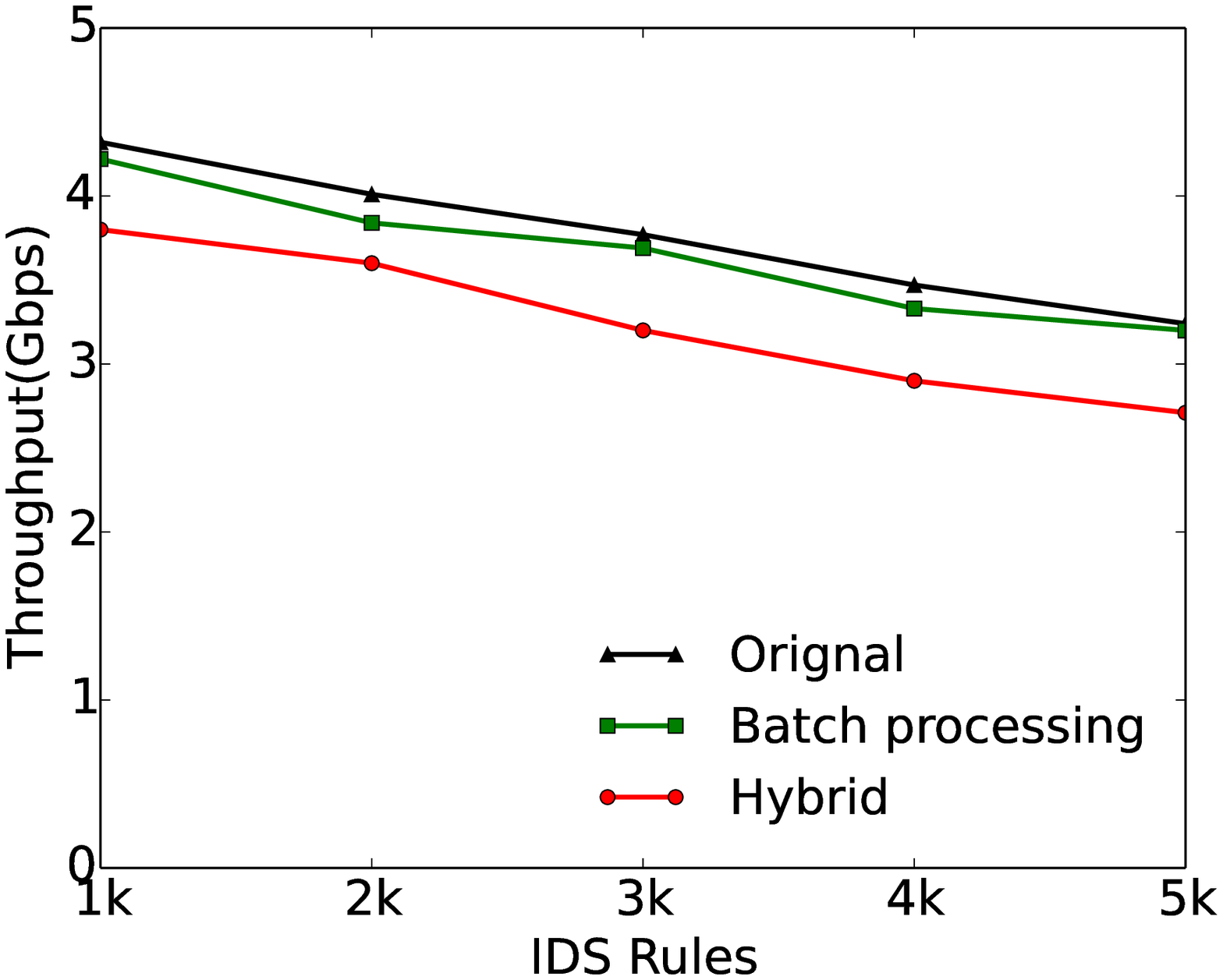}
\caption{Countermeasures Processing Delay (IDS and 600 bytes packets size)}
\end{minipage}
\begin{minipage}[t]{0.33\textwidth}
\centering
\includegraphics[width=6cm]{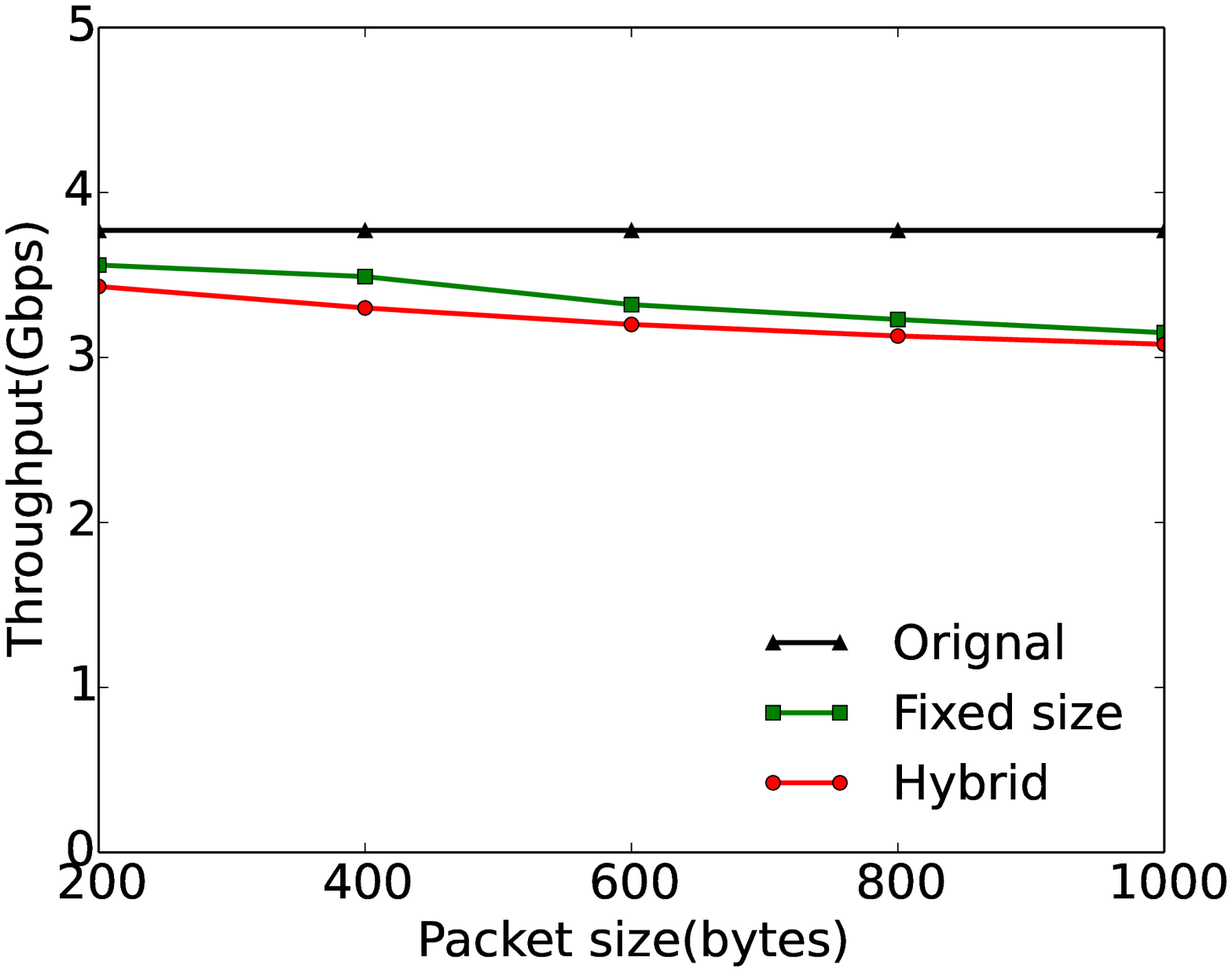}
\caption{Countermeasures Processing Delay (IDS with 3k rules)}
\end{minipage}
\end{figure*}

\subsection{Page Recognition Performance}
In this part, we will show the page recognition performance of the interface-based side channel attack on SGX-assisted NFV platforms. We will change the total amount of web pages contained in the web traffic, the percentage of tracked web pages to total web pages in different experiments, and the time interval between the collection of training data and test data. We show the accuracy and recall of identifying tracked web pages under each condition. Accuracy is the percentage of web pages identified accurately in all the identified web pages in our attack. It is calculated with the ratio of $W_{accurate\_identified}$ to $W_{total\_identified}$. $W_{accurate\_identified}$ refers to the amount of tracked web pages which are identified accurately in the attack. Once recognition is accurate only if current information buffer \textit{B} contains all the packets in corresponding profiling set \textit{P}. $W_{total\_identified}$ refers to the total amount of identified web pages by our page recognition algorithm. The recall is the percentage of web pages identified accurately of all the tracked web pages which should be identified in the web traffic. It is calculated with the ratio of $W_{accurate\_identified}$ to $W_{tracked}$. $W_{tracked}$ refers to the total amount of tracked web pages in traffic.
 
 To evaluate our attack performance under the circumstances of different web pages amount, we change the number of web pages which are visited simultaneously from 1k to 5k. Each test dataset corresponds to a training dataset whose amount of tracked web pages is half of the number of web pages contained in its corresponding test dataset. Figure 8 shows the experiment results. The accuracy and recall of our web page identifying algorithm can reach from 84.2\% to 97.1\% and from 76.6\% to 92.1\% respectively when the total amount of web pages in traffic varies from 1k to 5k. The accuracy drops when the number of web pages increases. This decline is caused by the common packets contained with different tracked web pages. When such packets appear, our page identifying algorithm will count it in every tracked web page. This will produce some false positive. However, the result shows that such false positive will remain less than 15.8\%. Some dynamic web pages change during the collection of training data and test data, thus some web pages don't be identified.
 
 To evaluate our attack performance under the conditions of different ratio of the amount of tracked web pages to the number of total web pages in traffic, we change this ratio from 25\% to 100\% in the traffic containing 3k web pages. As shown in Figure 9, the accuracy varies from 82\% to 87.6\%. There is no obvious changing tendency during the change of percentage of tracked web pages in the traffic. Thus, the performance of our tracked web page recognition algorithm is independent of the percentage of tracked web pages in the traffic.
 
 Most of the web pages in our dataset are dynamic web pages, they will change at the different time. The changing of the web pages will influence the recall of our web page recognition. To evaluate the performance of our attack at the different time, we change the time interval between the collection of training data set and test data set. Figure 10 shows the attack performance when the time interval varies from 10 min to 1 week. The accuracy ranges from 85.7\% to 88.9\% and the recall drops from 76.6\% to 43.7\% with the time interval varies from 10 min to 1 week. The decline of recall of our attack is caused by the dramatic change of tracked web pages. Thus, we need to refresh our training dataset every day to keep a stable recall.
 
\subsection{Packets Recognition Performance}
In this section, we will show the packets recognition performance of our interface-based side channel attack under the circumstances of the different total amount of web pages contained in web traffic and different percentage of tracked web pages to total web pages. We show the accuracy and recall of identifying packets which belong to the tracked web pages under each condition. Accuracy is the percentage of packets which actually belong to the identified website page in all the packets which are identified to belong to the identified website page according to our algorithm. It is calculated with the ratio of $P_{accurate\_identified}$ to $P_{total\_identified}$. $P_{accurate\_identified}$ refers to the number of packets which belongs to the identified web site page. A packet is identified accurately only if its IP address equal to the corresponding packets in the profiling set \textit{P}. $P_{total\_identified}$ refers to the number of packets which belong to the tracked website page according to our page recognition algorithm. The recall is the percentage of packets identified accurately in all the packets of tracked web site page. It is calculated with the ratio of $P_{accurate\_identified}$ to $P_{tracked}$. $P_{tracked}$ refers to the total amount of packets of tracked web site page. 

As shown in Figure 11, When the amount of the web pages varies from 1k to 5k, the accuracy of our packet recognition algorithm varies from 63.5\% to 68.3\% and the recall varies from 69.1\% to 71.7\%. As shown in Figure 12, when the ratio of the amount of the tracked web pages to the amount of the total web pages varies from 25\% to 100\%, the accuracy of our packet recognition algorithm varies from 64.3\% to 67.9\% and the recall varies from 68.4\% to 71.1\%. There is no obvious changing tendency of accuracy and recall under these two circumstances, which means the performance of our packets recognition algorithm is independent from the amount of the tracked web site pages and the percentage of tracked web pages in the traffic.


\subsection{Countermeasures Effectiveness and Overhead}
We mainly propose two countermeasures against privacy leakage problems in SGX-assisted applications. The batch operation mechanism will increase the processing delay because of the batch operation verification operations in the enclave. The fixing input data size mechanism will increase both processing delay and bandwidth cost. This is caused by the padding and un-padding operations in the enclave and increased copy operations produced by the packet padding. In this part, we first evaluate the countermeasures against our attack, then we will evaluate the overhead produced by these two countermeasures.

With our batch processing and fixing data size mechanisms, fine grained  side channel information cannot be collected precisely. Although the packet size can be collected in the network I/O modules in the OS. As shown in Figure 13. With only packet sizes which are equal to several fixed numbers, the accuracy of our attack is almost 0\% and the recall of our attack is almost 1\% when fixed packet size varies from 200 bytes to 1000 bytes. This shows that with these two countermeasures, privacy leakage in privacy-preserving NFV platform can be reduced to the volume which cannot make sense.

To measure the overhead of bandwidth produced by the fixing size of packet mechanism, we measure the bandwidth consumption of 5K web pages in our dataset when the fixed size of each packet varies. As shown in Figure 14, the original bandwidth of 5K web pages is 10.76 GB and the bandwidth consumption of 5K web pages varies from 11.26 GB to 12.94 GB when we change the fixed size of each packet from 200 bytes to 1000 byte. The overhead of the bandwidth varies from 4.6\% to 20\%. And with only 4.6\% additional bandwidth, the precision of our interface-based side channel attack can be reduced to nearly 0.

To measure the overhead of throughput produced by our two countermeasures. We first measure the throughput of SGX-assisted IDS with the number of rules from 1k to 5k. As shown in Figure 15, as the amount of IDS rules increase, the throughput of IDS without countermeasures decreases from 4.32 Gbps to 3.24 Gbps. This throughput decline is caused by the increase of IDS rules because IDS needs to spend time checking every rule. The throughput of IDS with batch processing mechanism decreases from 4.22 Gbps from 3.2Gbps. And the overhead of batch processing mechanism is mainly from the batch processing verification logic in the enclave, which is negligible. The throughput of IDS with two countermeasures decreases from 3.8 Gbps to 3.2Gbps. This decline mainly caused by the extra copy operations of padding data and the batch processing verification in the enclave. Then we measure the throughput of SGX-assisted IDS with 3k rules when the fixed packet size varies from 200 bytes to 1000 bytes. As shown in Figure 16, as the fixed size of packets increase, the throughput of IDS with fixing data size mechanism decreases from 3.56 Gbps to 3.15 Gbps. This throughput decline is caused by the increase of extra copy operation of padding data. The throughput of IDS with both mechanisms decreases from 3.43 Gbps to 3.08 Gbps, which is caused by extra copy operation of padding data and processing verification in the enclave.

\section{Related Work}

\subsection{Side Channel Attacks in SGX}
Intel SGX can ensure the integrity and confidentiality of data processing in the enclave. However, it suffers various kinds of side channel attacks. Several works have exploited different side channel to attack SGX.
Xu et al.\cite{xu2015controlled} and Shinde et al.\cite{shinde2015preventing} demonstrate the controlled channel attack which intentionally manipulates the page table which is managed by the untrusted OS to infer the secret inside the enclave. Lee et al.\cite{lee2017inferring} illustrate the branch shadowing attack by constructing codes which share the Branch Target Buffer (BTB) with enclave codes and observing Last Branch Record (LBR) to infer the fine-grained control flow of enclave codes. Ferdinand et al.\cite{brasser2017software}, Moghimi et al.\cite{moghimi2017cachezoom} and Schwarz et al.\cite{schwarz2017malware} demonstrate the cache-based side channel attack on SGX to infer the AES keys or RSA keys in the enclave.

\subsection{Encrypted Traffic Analysis Attack}
Hintz et al.\cite{hintz2002fingerprinting} use the file sizes as features to identify individual websites when the server is not known. Sun et al.\cite{sun2002statistical} use Jaccard's coefficient to match the websites with slightly varying sizes to pre-collected traffic patterns. However, now HTTP makes use of persistent connections and pipelining to meet the requirement of high performance, it is no longer possible to satisfy the assumption that distinguishing between single web object requests in the above works. Bissias et al.\cite{bissias2005privacy} perform website fingerprinting base on IP packet sized and inter-packet arrival times instead of web object sizes. Liberatore et al.\cite{liberatore2006inferring} compared the effectiveness of the na ̈ıve Bayes classifier and Jaccard’s coefficient on SSH-protected channels. Lu et al.\cite{lu2010website} showed that website fingerprinting can be improved by considering information about packet ordering. However, all of the above works assume that the websites are visited one by one. Such an assumption is far from real which multiple pages are visited simultaneously. The recognition methods proposed above will not work in multi-page scenarios.
 
\section{Conclusions}
Intel SGX has been considered as one of the most promising hardware-based technology to realize the TEE. However, it suffers from various side channel attacks. In this paper, we explored a new kind of side channel attack against Intel SGX, called interface-based side channel attack, which can infer secret information contained in the enclave input data. We also propose some countermeasures to reduce the exposure of interface-based side channel information during program executing.


\bibliographystyle{ACM-Reference-Format}
\bibliography{sample-bibliography}

\end{document}